# Asymmetric Frequency Conversion and Violation of Kirchhoff's Law of Thermal Radiation in Phased Floquet Metasurfaces

Yao-Ting Wang,[1,2] Emanuele Galiffi,[2,3] Sander A. Mann,[2,4] Mauro Antezza,[5,6] and Andrea Alù[2,7,8]

[1]Department of Photonics, National Sun Yat-sen University, Kaohsiung, 804201, Taiwan

[2]Photonics Initiative, Advanced Science Research Center, City University of New York, New York, New York 10031, USA

[3]Chandra Department of Electrical and Computer Engineering, The University of Texas at Austin, 2501 Speedway, Austin, TX, USA

[4]Institute of Physics, University of Amsterdam, Amsterdam, The Netherlands

[5]Laboratoire Charles Coulomb (L2C), UMR 5221 CNRS-Université de Montpellier, F-34095 Montpellier, France

[6]Institut Universitaire de France, 1 rue Descartes, Paris Cedex 05 F-75231, France

[7]Physics Program, Graduate Center, City University of New York, New York, New York 10026, USA

[8]Department of Electrical Engineering, City College of New York, New York, New York 10031, USA

*Correspondence: ytwang@mail.nsysu.edu.tw (Y.-T.W.), aalu@gc.cuny.edu (A.A.)

[9]Lead contact: Andrea Alù, aalu@gc.cuny.edu

## SUMMARY

**We study a system of three closely spaced conducting sheets, whose thermal emission largely violates Kirchhoff's law through nonreciprocal frequency conversion enabled by Floquet phenomena. We leverage surface mode transitions controlled by temporal modulation of one of the metasurfaces, resulting in largely asymmetric frequency conversion. Our findings broaden the capabilities of Floquet metasurfaces for wave conversion and energy manipulation, paving the way towards magnet-free nonreciprocal photonic devices with opportunities for thermal management and efficient photonic refrigeration.**

## INTRODUCTION

The manipulation of electromagnetic waves through engineered structures has been a cornerstone of modern photonics, with significant advances driven by the development of metamaterials.[1,2] Conventional approaches rely on static structures carefully tailored in space, limiting their functionality and adaptability. More recently, Floquet metamaterials, whose material or structural parameters are modulated periodically in time,[3,4] have emerged as a dual platform overcoming some of the limitations of static metamaterials, and yielding efficient frequency conversion[5,6] and nonreciprocal wave propagation.[7–9] In particular, nonreciprocity is fundamental for many technologies, from radio-frequencies to optics, both for classical and quantum applications that require gyrators, isolators[10] and circulators.[11] In addition to nonreciprocity, time-modulated systems offer dynamic control over the wave properties, enabling functionalities and phenomena unattainable in passive, time-invariant structures, such as artificial magnetic fields for photons,[12] temporal wavepacket aiming,[13] photonic refrigeration with time-varying media,[14,15] photonic time crystals,[16–20] reflection/transmission on a time interface,[21–27] among others.[28–37] Beyond applications for magnet-free nonreciprocal components, the exploration of time-modulated systems has also led to the rapid development of photonic time crystals, which exhibit rich wave physics and distinctive momentum-bandgap phenomena.[38]

Floquet metasurfaces have emerged as a cutting-edge frontier in this context. They can support Floquet phenomena via boundary conditions periodically modulated in time, yielding anomalous reflection and transmission,[39] polarization control[40] and dynamic beam shaping.[41] Recent advances in fabrication techniques and material science have propelled the development of Floquet metasurfaces, making them increasingly feasible for practical applications like beam steering,[42] propagating-to-surface-wave conversion[43] and selective surface wave excitation.[44]

In the following, we explore phased Floquet metasurfaces to realize asymmetric frequency conversion, yielding large violation of Kirchhoff's law of thermal radiation driven by Floquet phenomena. Unlike previous designs based on coupled-ring resonators[45] or spatio-temporal traveling wave bias,[46] our uniform stacked Drude-sheet structure breaks reciprocity through purely temporal multi-tone modulation, eliminating the need for spatial modulation while preserving spatial symmetry. Our configuration uses a fast modulation tone to efficiently couple free-space radiation into bound surface modes. By introducing two other modulation tones with a proper phase relation among them, we generate nonreciprocal frequency conversion for free-space radiation. In addition, by exclusively selecting single-mode loss in the unmodulated bottom sheet, our platform offers a distinct route to achieving large violation of Kirchhoff's law of thermal emission through a purely time-modulated system. To realize this concept, we first characterize the surface----wave modes supported by the three--sheet structure. We then introduce the phased Floquet modulation and show how it enables asymmetric frequency conversion. Finally, we demonstrate how this mechanism leads to a strong violation of Kirchhoff's law, by showing how surface waves in this structure can be selectively excited, frequency converted, outcoupled and/or absorbed beyond the constraints of time-invariant systems.

## RESULTS

Our geometry (Figure 1) consists of three closely spaced conducting sheets with identical passive properties, enabling three equally spaced surface modes. The top sheet, loaded by electrically tunable components, is driven with a periodic modulation pattern creating tailored wave interference to induce selective frequency conversion and full violation of Kirchhoff's law. The induced surface current density $J_x$ on the top sheet is governed by the Drude dynamic equation

$$\dot{J}_x + \gamma J_x = W_D\left(t\right) E_x, \tag{1}$$

where $\gamma$ is the loss factor, $W_D\left(t\right) = W_{D0}\left[1 + \sum_{j=1}^{3} 2\alpha_j \cos\left(\Omega_j t + \phi_j\right)\right]$ is the Drude weight. The index $j = 1, 2, 3$ denotes the three distinct tones of the multi-frequency modulation signal. $W_{D0}$, $\alpha_j$, $\Omega_j$, and $\phi_j$ describe the static response and the modulation coefficients of the top surface,[43] where $j$ runs from 1 to 3 to account for each modulation component. The modulation signal is composed of three driving frequencies and it is applied to the top metasurface, serving distinct physical roles: the modulation amplitudes govern the overall coupling strength between the resonant modes of the system; the first and second modulation frequencies, oscillating at Ω and 2Ω, mediate inter-band coupling between the equally-spaced surface bands; the third modulation frequency, characterized by a rapidly oscillating component, enables the interaction of vacuum modes above the light cone into surface-bound states. $E_x$ is the transverse component of the electric field for transverse-magnetic (TM) polarized excitation ($H_y$, $E_x$, $E_z$). The middle and bottom layers are static conducting sheets, whose Drude weight is also $W_{D0}$, and $d_A$, $d_B$, $d_C$ are the distance between Floquet sheets and a perfect electric conductor (PEC) back-reflector. This PEC is valid in the GHz–THz regime where our design is targeted, as the high conductivity of metals at these frequencies ensures negligible skin depth. Lastly, the temporal modulation phases associated with each component introduce controlled phase shifts during mode conversion, giving rise to nonreciprocal interference phenomena, remarkably without the need to break spatial symmetry.

The expression for the TM-polarized in-plane electric field upon multiple scattering from the metasurfaces is given by

$$E_{x,j}\left(x,z,t\right)=\sum_{n}\left(a_{jn}e^{ik_{z,jn}z}+b_{jn}e^{-ik_{z,jn}z}\right)e^{i\left(k_{x}x-\omega_{n}t\right)}, \tag{2}$$

where $a_{jn}\left(b_{jn}\right)$ is the complex amplitude of the forward (backward) propagating Floquet harmonic, $k_x$ is the in-plane wavevector, $k_{z,jn}=\sqrt{\varepsilon_j\,\omega_n^2/c^2-k_x^2}$ is the wavevector component perpendicular to the modulated sheets. The harmonic frequencies are expressed as $\omega_n=\omega+n\Omega$, where $\Omega$ is the modulation frequency and *n* indicates the Floquet harmonic order. Without loss of generality, we assume that the dielectric constant between sheets is $\varepsilon_j=\varepsilon_0$ for all regions *j = 0, A, B, C*, implying a uniform air environment between the layers and the ground plane reflector placed behind the last conducting sheet. The corresponding in-plane magnetic field components are

$$H_{y,j}\left(x,z,t\right)=\sum_{n}\frac{\varepsilon_0\varepsilon_j\omega_n}{k_{z,jn}}\left(a_{jn}e^{ik_{z,jn}z}-b_{jn}e^{-ik_{z,jn}z}\right)e^{i\left(k_{x}x-\omega_{n}t\right)}. \tag{3}$$

In the absence of temporal modulation, the three coupled layers support three surface modes, as shown in Figure 2a by the colored dispersion curves. We choose the normalized distances $\tilde{d}_A=1$, $\tilde{d}_B=2.7$, $\tilde{d}_C=10$, where $\tilde{d}_j=d_j\left(W_{D0}Z_0/c\right)$, to induce two equi-frequency gaps between surface modes at normalized wavevector $\tilde{k}_x=0.6$, where $\tilde{k}_x=k_x/\left(W_{D0}Z_0/c\right)$. The eigenfields of each surface wave are shown in the inset of Figure 2a. The eigenfield for $\left|s_3\right\rangle$ presents an anti-symmetric mode primarily located in the top and middle layer, while the eigenfield for $\left|s_1\right\rangle$ corresponds to a symmetric mode distributed across all three sheets. Finally, the eigenfield for $\left|s_2\right\rangle$ exhibits a field symmetry similar to $\left|s_3\right\rangle$, but its field strength is localized near the bottom layer.

The Floquet modulation of the top surface aims at converting incoming waves from free-space at oblique incidence into surface waves through frequency conversion in an asymmetric fashion, while preserving transverse momentum. In Eq. (1), the modulation frequencies are denoted as $\Omega_1=\tilde{\Omega}$, $\Omega_2=2\tilde{\Omega}$, and $\Omega_3=9\tilde{\Omega}$, in which $\tilde{\Omega}=\Omega/W_{D0}Z_0=0.11$ is the normalized frequency. We choose the value of $\Omega_3$ to be significantly larger than the maximum frequency difference between any two surface modes to prevent interference between generated surface waves. To determine the amplitudes in Eqs. (2) and (3), we utilize the continuity conditions for the transverse electric $E_{x,j}$ and magnetic $H_{y,j}$ fields associated with the surface current $J_x(x,t) = \sum_n J_n e^{i(k_x x - \omega_n t)}$, where

$$J_n = \left(\frac{W_{D0}}{\gamma - i\omega_n}\right)\Big[E_{x,jn} + \alpha_1\left(E_{x,n+1}e^{i\phi_1} + E_{x,n-1}e^{-i\phi_1}\right) + \alpha_2\left(E_{x,n+2}e^{i\phi_2} + E_{x,n-2}e^{-i\phi_2}\right) + \alpha_3\left(E_{x,n+9}e^{i\phi_3} + E_{x,n-9}e^{-i\phi_3}\right)\Big]. \tag{4}$$

Applying the boundary conditions at the three surfaces leads to a $2n \times 2n$ transfer matrix for the Floquet system, which can be solved using matrix inversion.[44]

As shown in Figure 2b, when the frequency detuning between the incident wave and a surface mode matches the fast modulation frequency $\Omega_3$, the corresponding surface wave can be excited from free space. Notably, because of the degeneracy of the modulation signals, by varying the modulation phases of the three signals we can control the coupling among modes. In particular, the proper choice of phases can break symmetry in mode conversion, and by flipping the sign of the modulation phases we obtain a time-reversal operation, reversing the symmetry breaking. To aid intuition, this mechanism can be visualized as three coupled surface modes connected by Floquet-induced transition pathways, where each mode couples to its neighbors through frequency-conversion processes. This conceptual representation of mode-to-mode coupling clarifies how wave interference enables the inter-modal circulator illustrated in the bottom part of Figure 2b, where the relative phases dictate the direction of the coupling sequence. This physical circulation is directly manifested in the reflectance spectra of Figure 3, where the phase-controlled transitions follow distinct color-coded pathways: blue denotes reflection without conversion, red corresponds to up-conversion to the next higher mode, and yellow (or green/purple in the opposite-phase case) indicates down-conversion to the next lower mode.

In the following results, we choose modulation phases $(\phi_1, \phi_2, \phi_3) = (0, \pm 0.1\pi, 0)$, which are optimized to yield such a response. In particular, Figure 3a shows the calculated reflectance spectra for the $0^{th}$, $\pm 1^{st}$, and $\pm 2^{nd}$ frequency channels for the two signs of $\phi_2$ and optimal values for the modulation amplitudes, as indicated in the figure, for a fast modulation frequency $\tilde{\Omega}_3 = 0.99$ that enables coupling of surface waves to free-space radiation for the selected incidence angle. By choosing $\phi_2 = +0.1\pi$ (Figure 3a-b), an incident wave in the lowest resonant state $|i_1\rangle$ first couples to surface mode $|s_1\rangle$ and then it is converted to $|s_3\rangle$, which resonantly radiates back into the continuum state $|i_3\rangle$ with a net frequency shift of $2\Omega$. This sequence follows the handedness illustrated in Figure 2b, producing preferential up-conversion for the lowest mode. Conversely, illumination via the continuum states $|i_2\rangle$ and $|i_3\rangle$ excites surface modes $|s_2\rangle$ and $|s_3\rangle$, respectively converted into $|s_1\rangle$ and $|s_2\rangle$ surface modes by the modulation, and then re-radiated back with a net frequency shift of $-\Omega$. By contrast, setting the modulation phase to $\phi_2 = -0.1\pi$ (Figure 3c-d) time-reverses the modal conversion process, leading to frequency shifts $-2\Omega$ and $+\Omega$, since the propagating state $|i_3\rangle$ is converted into $|i_1\rangle$, while $|i_2\rangle$ and $|i_3\rangle$ are converted into $|i_1\rangle$ and $|i_2\rangle$, respectively. In the figure, transfer matrix calculations are compared to COMSOL simulation results, validating the described nonreciprocal surface modal transition.

Specifically, Figure 3a and the blow-ups in Figure 3b show that, for $\phi_2 = 0.1\pi$, if the incident wave impinges on the structure at the lowest frequency $\tilde{\omega} = 1.302$ ($|i_1\rangle$), its own reflection (blue) is minimal, while a $+2\tilde{\Omega}$ frequency up-conversion (red) to $|i_3\rangle$ occurs via surface modal conversion $|s_1\rangle \rightarrow |s_3\rangle$. In a similar manner, for frequencies $\tilde{\omega} = 1.412$ ($|i_2\rangle$) and $\tilde{\omega} = 1.522$ ($|i_3\rangle$), a $-\Omega$ frequency down-conversion (yellow) is observed, respectively mediated by the surface-wave transitions $|s_2\rangle \rightarrow |s_1\rangle$ and $|s_3\rangle \rightarrow |s_2\rangle$, as highlighted in Figure 3b. Conversely, Figure 3c and 3d demonstrate how flipping the sign of $\phi_2 = -0.1\pi$, the surface modal transitions follow opposite handedness: upon excitation at $\tilde{\omega} = 1.302$ ($|i_1\rangle$) and $\tilde{\omega} = 1.412$ ($|i_2\rangle$), the system supports $+\tilde{\Omega}$ transitions (green) to $|i_2\rangle$ and $|i_3\rangle$, via $|s_1\rangle \rightarrow |s_2\rangle$ and $|s_2\rangle \rightarrow |s_3\rangle$, respectively. For an incident frequency $\tilde{\omega} = 1.512$, a $-2\tilde{\Omega}$ transition (purple) occurs $|i_3\rangle \rightarrow |i_1\rangle$, mediated by the modal transition $|s_3\rangle \rightarrow |s_1\rangle$. These results clearly highlight the nonreciprocal modal interactions and intermodal circulation of the system, which reproduce the scheme in Figure 2b for free-space radiation. Remarkably, this is obtained by simply modulating uniformly in space one of the three conducting layers. The asymmetric conversion mechanism is robust against detuning of the modulation phase and of the interlayer spacing, as detailed in Supplementary Note S1 (Figures S1 and S2).

The described interplay between time modulation and modal asymmetry in our Floquet system enables not only asymmetric frequency conversion, but it is also ideally suited to access large violation of Kirchhoff's law of thermal radiation. In conventional, time-invariant systems in thermal equilibrium, the emissivity at a given frequency, direction and polarization is expected to be equal to its absorptivity. However, in nonreciprocal structures, this relation no longer has to hold, enabling unique thermal features of great interest for energy applications.[14,15]

In our system, large violations of emission-absorption symmetry can be accessed by introducing losses at critical coupling. We introduce an absorption rate $\gamma = 3\times10^{-3}\omega_p$ in the bottom (unmodulated) sheet (see schematic inset in Figure 4), where eigenmode $|s_2\rangle$ is predominantly localized, as shown in Figure 2a. The damping rate $\gamma = 3\times10^{-3}\omega_p$ is chosen to represent the loss rate in realistic material systems and electronic components, rather than an optimized parameter. Numerical analysis confirms that the resulting violation of Kirchhoff's law remains robust across a range of damping values, which ensure experimental feasibility. This localization ensures that excitations involving $|s_2\rangle$ incur significant dissipation, while other excitations are minimally impacted by losses. The absorption of the system can be evaluated as the power dissipated by the bottom sheet

$$A = \sum_m A_m = \sum_m \frac{1}{2}\mathrm{Re}\left\{\frac{W_{D0}}{\gamma - i\omega_m}\right\}\left|E_{x,m}\right|^2, \tag{5}$$

and the emitted power can be evaluated as[47]

$$\mathcal{E}(\omega) = \sum_m \frac{\Theta(\omega_m, T)}{\Theta(\omega, T)}\left(\frac{\omega}{\omega_m}\right)^2 A_m^{\mathcal{T}}, \tag{6}$$

where $\mathcal{E}$ is the total emissivity (in power basis) at frequency $\omega$, the summation is over the partial harmonics, $\omega_m = \omega + m\Omega$, $\Theta(\omega,T) = \hbar\omega/[\exp(\hbar\omega/k_B T)-1]$ is the mean thermal energy of a harmonic oscillator and $A_m^{\mathcal{T}}$ is the absorption (in power basis) in each harmonic in the adjoint system (with time-reversed bias). Absorption and emission are defined with respect to the incident and outgoing frequency, and the harmonics are defined with respect to those frequencies. Equation 5 stems from the fact that absorption and emission are time-reversed processes for each harmonic, which is the origin of the relation in Equation 6.[47] The frequency prefactor is a consequence of the Manley–Rowe relations, stating that the power is increased if e.g. a photon is generated in a given harmonic but upconverted before being emitted. At low frequencies or high temperatures we can approximate thermal fluctuations as Johnson-Nyquist noise, which yields $\Theta(\omega,T) = k_B T$, simplifying Equation 6. In the following, we assume to be in the regime dominated by classical noise. Note that in the limit of closely spaced harmonics with approximately equal thermal photon occupations, and in the absence of parametric amplification such that photon flux is conserved, Equation 6 reduces to the scattering-matrix expression derived by Ghanekar et al.[46]

Figure 4 shows absorptivity and emissivity for the two modulation schemes across the entire frequency range under analysis, while the insets zoom into the frequency ranges around the resonances, where strong violation of Kirchhoff's law can be observed. First, we notice that the emissivity is generally significantly larger than the absorptivity across all resonances and for both modulation schemes, a consequence of the fact that incident light is downconverted to be absorbed, but upconverted to be emitted: due to the fast modulation frequency $\Omega_3 = 9\tilde{\Omega}$, the absorption and emission arise most prominently at harmonic indices from -7 to -11. At lower frequencies there is a larger spectral density of thermal fluctuations, resulting in larger emissivity and photonic refrigeration.[14] For instance, near $\tilde{\omega} = 1.30$, around the lower resonance frequency, the $+2\tilde{\Omega}$ transition occurs so that the absorption with $\phi_2 = 0.1\pi$ and the emission with $\phi_2 = -0.1\pi$ avoid $|s_2\rangle$, yielding large spectral responses, while for the $+\tilde{\Omega}$ route absorption with $\phi_2 = -0.1\pi$ and emission with $\phi_2 = 0.1\pi$ traverses $|s_2\rangle$, increasing dissipation and modifying the emitted spectrum. A similar but phase-reversed asymmetry is observed near $\tilde{\omega} = 1.52$, involving the $-2\tilde{\Omega}$ and $-\tilde{\Omega}$ transitions, which originate due to the relationship between emissivity and absorptivity in time-reversed pairs of the time-modulated system as shown in Equation 5.

## DISCUSSION

The pronounced emissivity–absorptivity mismatch near each resonance arises because the phased Floquet modulation enforces a directional sequence of inter-modal conversions: absorption (from a radiative channel) and thermal emission (from the lossy mode) traverse different, phase-selected pathways through the selectively lossy mode. This strong difference between emissivity and absorptivity is a direct macroscopic manifestation of the intermodal circulation established in the structure. Beyond the large imbalance between absorption and emission due to the frequency conversion, the strong nonreciprocal response of the system and intermodal circulation impart sharply different spectral responses in emissivity and absorptivity around each resonance, confirming the large violation of Kirchhoff's thermal radiation. The temporal modulation acts as an external work reservoir, supplying or receiving energy for each photon converted between harmonics as encoded by the Manley–Rowe factors in Equation 6. The thermodynamic performance of this class of devices has been analyzed for time-modulated thermal emission by Buddhiraju et al.[14]

In this work, we investigated asymmetric frequency conversion and broken absorption-emission symmetry in phased Floquet metasurfaces. Our analysis demonstrates that proper forms of multi-frequency time-modulation introducing broken symmetries in time can effectively induce handedness and circulation in surface modal transitions, which in turn lead to selective frequency conversion mediated by surface waves. Spatially selective absorption enables large violations of Kirchhoff's thermal radiation law, owing to the interplay between the nonreciprocal nature of the modulated system and the presence of material loss selective to specific modes. Our work provides a foundational understanding of advanced nonreciprocal systems based on time modulation, highlighting opportunities offered by Floquet metasurfaces for wave manipulation and energy control, and paving the way for applications in active nonreciprocal photonic devices. Beyond its theoretical significance, the proposed phased Floquet metasurface offers a viable path toward experimental realization of these phenomena across different spectral regimes.

In the microwave regime, our design is readily accessible using established technologies. Specifically, the time-modulated conductive sheets can be implemented by graphene-based metasurfaces with realistic mobilities[46] or varactor-loaded metallic networks on printed circuit boards.[48,49] These platforms offer reliable, magnet-free routes to realize the necessary fast multi-tone modulation and phase control. Conversely, scaling this mechanism to optical frequencies using active materials like Indium Tin Oxide (ITO) requires sophisticated all-optical modulation schemes.[35,50] Achieving the required modulation amplitudes for significant optical nonreciprocity would require further structuring and optimization of the pumping setup, at the cutting edge of current experimental capabilities. Therefore, while our theoretical framework is broadly generalizable, we envision experimental verifications to be focused primarily in the GHz–THz regime, which can serve as a robust foundation to motivate future advances in ultrafast active optical materials. Looking forward, we anticipate that this platform may be extended to more complex wave manipulation regime and to thermal management devices.

## METHODS

### *Transfer-matrix formalism*

The electromagnetic response of the three-sheet Floquet structure was obtained from the Floquet transfer-matrix formalism defined by Equations 1-4. The fields in each region were expanded in Floquet harmonics and truncated at $|n| = 13$. Imposing the continuity conditions for the transverse electric and magnetic fields at the three sheets yields a $2n \times 2n$ linear system, which was solved by matrix inversion.

### *Full-wave simulations*

Full-wave validation was carried out with the RF Module of COMSOL Multiphysics. 27 frequency-domain electromagnetic interfaces, one per harmonic channel with n running from -13 to 13, were coupled through the surface current density prescribed on the modulated sheet. Floquet periodic conditions were applied on the lateral boundaries, perfectly matched layers truncated the domain above the structure, and a perfect electric conductor terminated it below. The reflectance of each harmonic channel was computed from the ratio of the reflected to the incident time-averaged Poynting vectors and weighted by the Manley–Rowe frequency factor. A schematic of the simulation domain is shown in Figure S3.

### *Absorption and emission spectra*

The absorptivity was evaluated as the power dissipated in the lossy bottom sheet, Equation 5. The emissivity was obtained from Equation 6, which relates the emission of the system to the absorption of the adjoint system with time-reversed bias, evaluated in the classical Johnson–Nyquist noise limit.

Further details regarding the methods can be found in Supplementary Note S2.

## RESOURCE AVAILABILITY

### *Lead contact*

Requests for further information and resources should be directed to and will be fulfilled by the lead contact, Andrea Alù, aalu@gc.cuny.edu.

### *Materials availability*

This study did not generate new unique reagents.

### *Data and code availability*

- All data reported in this paper will be shared by the lead contact upon request.
- This paper does not report original code.
- Any additional information required to reanalyze the data reported in this paper is available from the lead contact upon request.

## ACKNOWLEDGMENTS

This work was supported by the Air Force Office of Scientific Research and the Simons Foundation. Y.-T.W. acknowledges financial support from the National Science and Technology Council under Grant No. NSTC 113-2112-M-110-007-MY3. The authors declare no conflicts of interest.

## AUTHOR CONTRIBUTIONS

Conceptualization, Y.-T.W., E.G., S.A.M., and A.A.; methodology, Y.-T.W., E.G., and S.A.M.; software and investigation, Y.-T.W.; formal analysis, Y.-T.W., S.A.M., and M.A.; writing - original draft, Y.-T.W.; writing - review & editing, Y.-T.W., E.G., S.A.M., M.A., and A.A.; supervision and funding acquisition, A.A.

## DECLARATION OF INTERESTS

The authors declare no competing interests.

## DECLARATION OF GENERATIVE AI AND AI-ASSISTED TECHNOLOGIES IN THE WRITING PROCESS

During the preparation of this work, the author(s) used Anthropic Claude in order to avoid grammatical and typographical errors. After using this tool or service, the author(s) reviewed and edited the content as needed and take(s) full responsibility for the content of the publication.

## SUPPLEMENTAL INFORMATION

Document S1. Supplementary Notes S1 and S2, and Supplementary Figures S1–S3.

## FIGURE TITLES AND LEGENDS

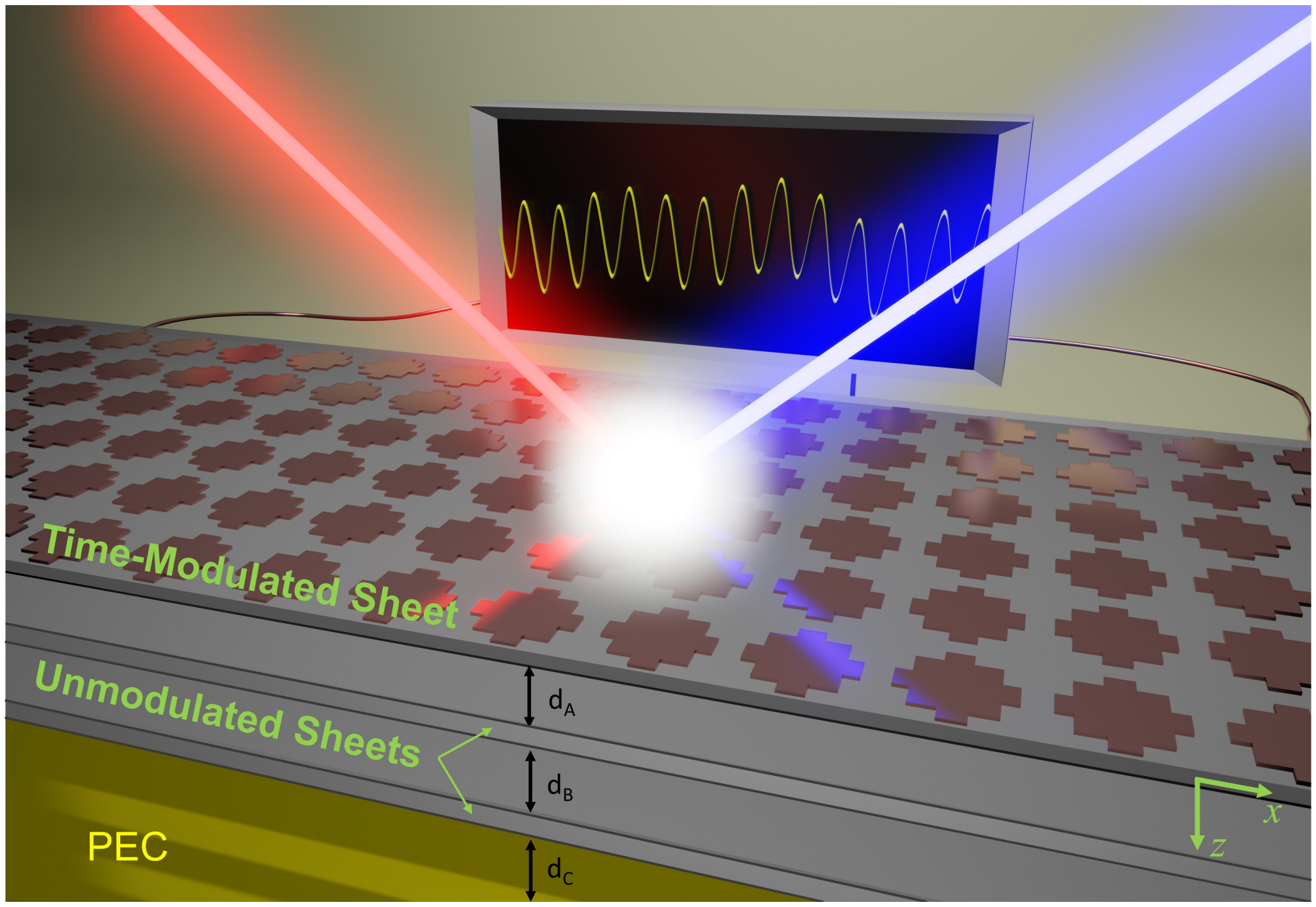

*Figure 1. Schematic of the phased Floquet metasurface.*

Schematic representation of the proposed geometry, with incident and reflected waves represented by red and blue beams. The top metasurface Drude weight is modulated in time. Beneath the Floquet metasurface are two unmodulated sheets and a ground plane. The distance between neighboring layers and the reflector are denoted as $d_A$, $d_B$ and $d_C$. The sheet modulation is illustrated by the waveform on the screen in the background, with parameters $(\alpha_1,\alpha_2,\alpha_3)=(0.05,0.05,0.2)$ and $(\Omega_1,\Omega_2,\Omega_3)=(1,2,9)\tilde{\Omega}$.

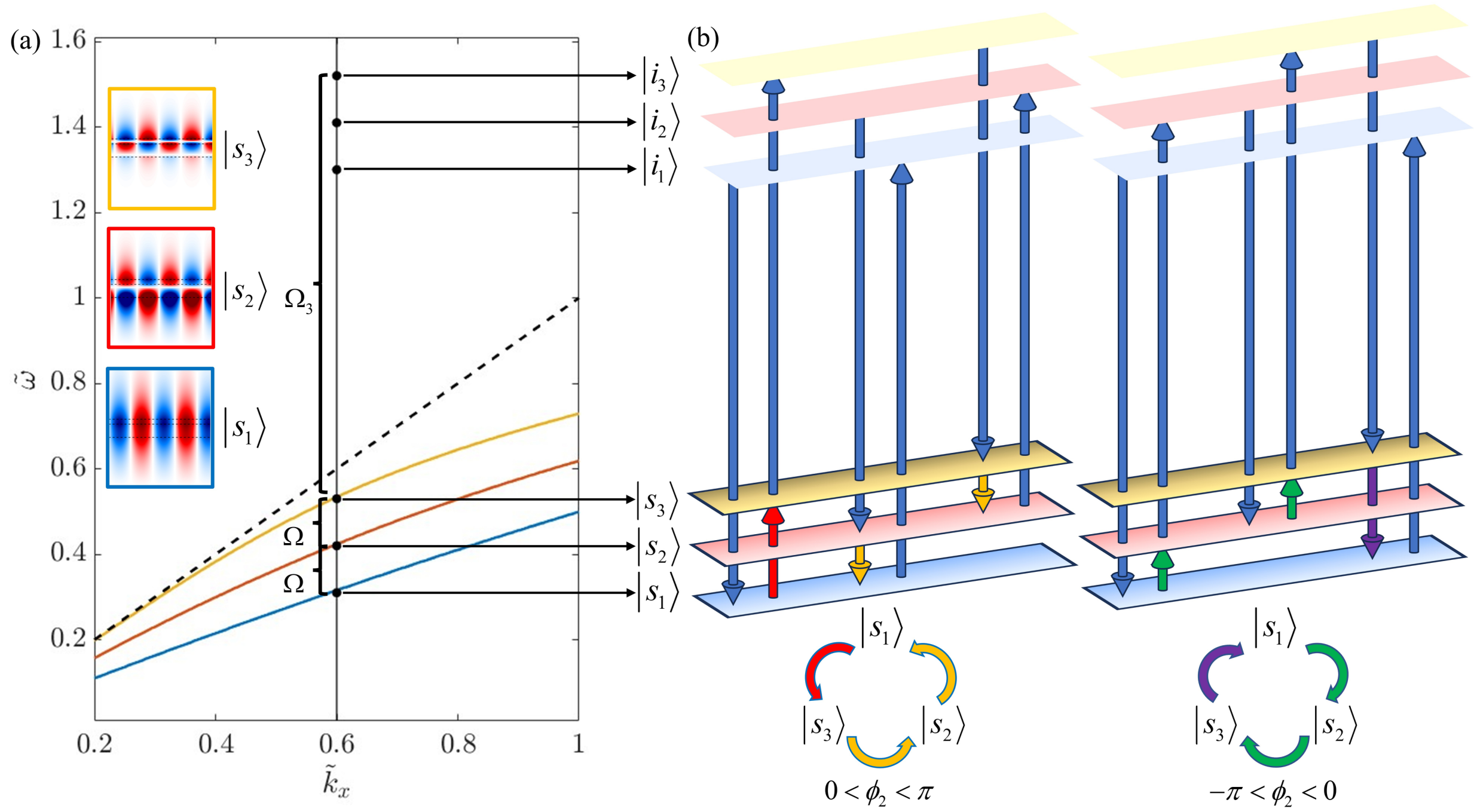

*Figure 2. Surface-mode dispersion and phase-controlled inter-modal circulation.*

(a) Dispersion relation and mode profiles when the top metasurface is modulated. The dispersion curves for the three modes $|s_1\rangle$, $|s_2\rangle$ and $|s_3\rangle$ are equally spaced at the normalized wavevector $\tilde{k}_x$. The insets depict the eigenfield profiles. (b) Schematic of the mode coupling for different modulation phases. The left diagram shows the coupling for a positive modulation phase $0<\phi_2<\pi$, while the right diagram shows the coupling for a negative modulation phase $-\pi<\phi_2<0$. Arrows indicate the coupling handedness, demonstrating phase-dependent interaction among modes.

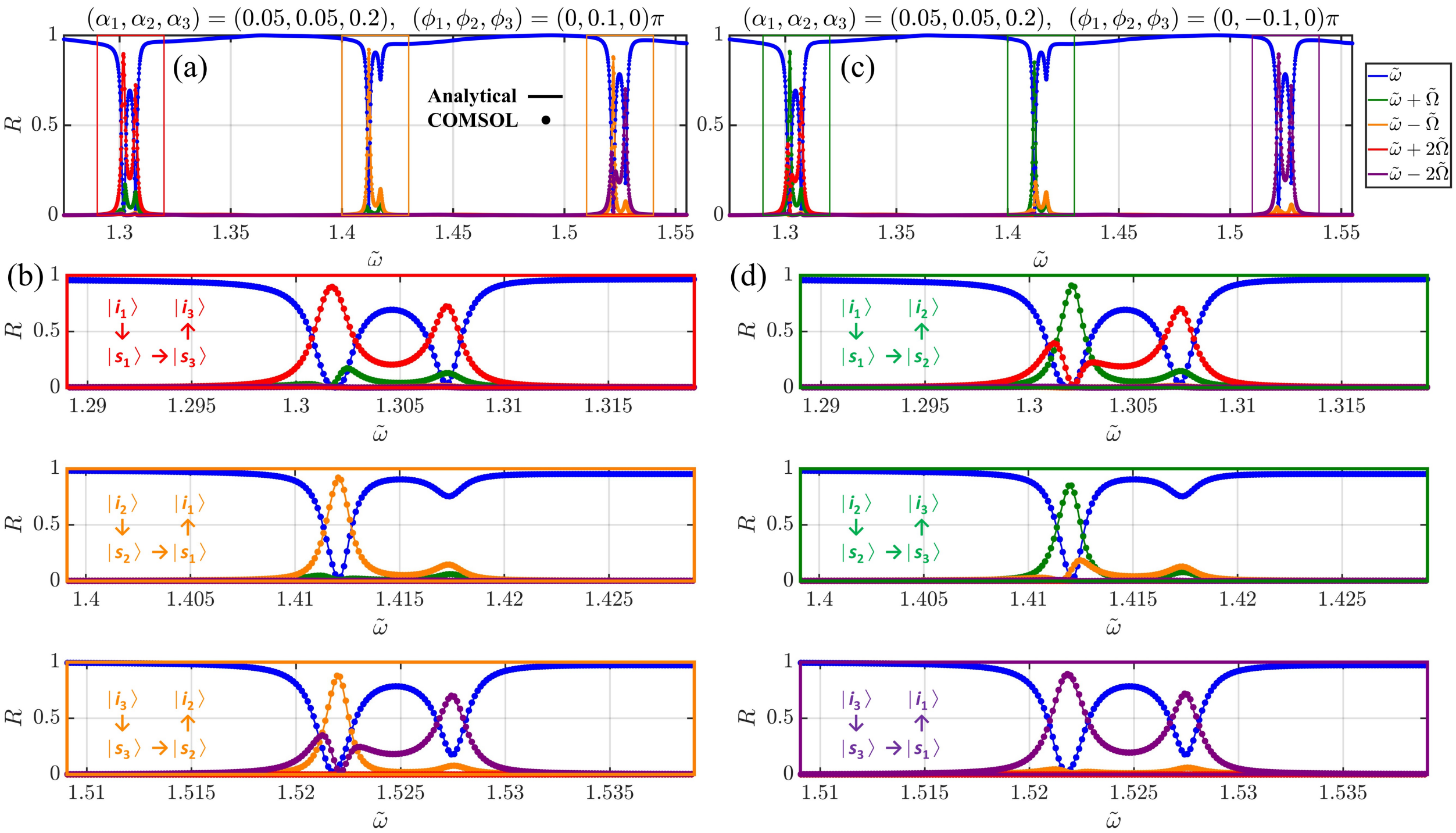

*Figure 3. Reflectance spectra showing phase-controlled modal transitions in the phased Floquet metasurface*

Colors indicate distinct frequency-conversion pathways between the three surface modes (blue: reflection; red: up-conversion; yellow/green/purple: down-conversion). In both cases, the modulation strengths are set to $(\alpha_1,\alpha_2,\alpha_3)=(0.05,0.05,0.2)$. Panels (a,b) correspond to modulation phases $(\phi_1,\phi_2,\phi_3)=(0,+0.1\pi,0)$, while panels (c,d) use phases of opposite sign, $(\phi_1,\phi_2,\phi_3)=(0,-0.1\pi,0)$. The comparison between the two cases demonstrates a full, phase-controlled reversal of the modal-conversion direction and the exchange of up/down-conversion channels. Solid lines represent analytical transfer matrix calculations, and dots show COMSOL simulation results, validating the nonreciprocal surface modal transitions. See also Figures S1–S3.

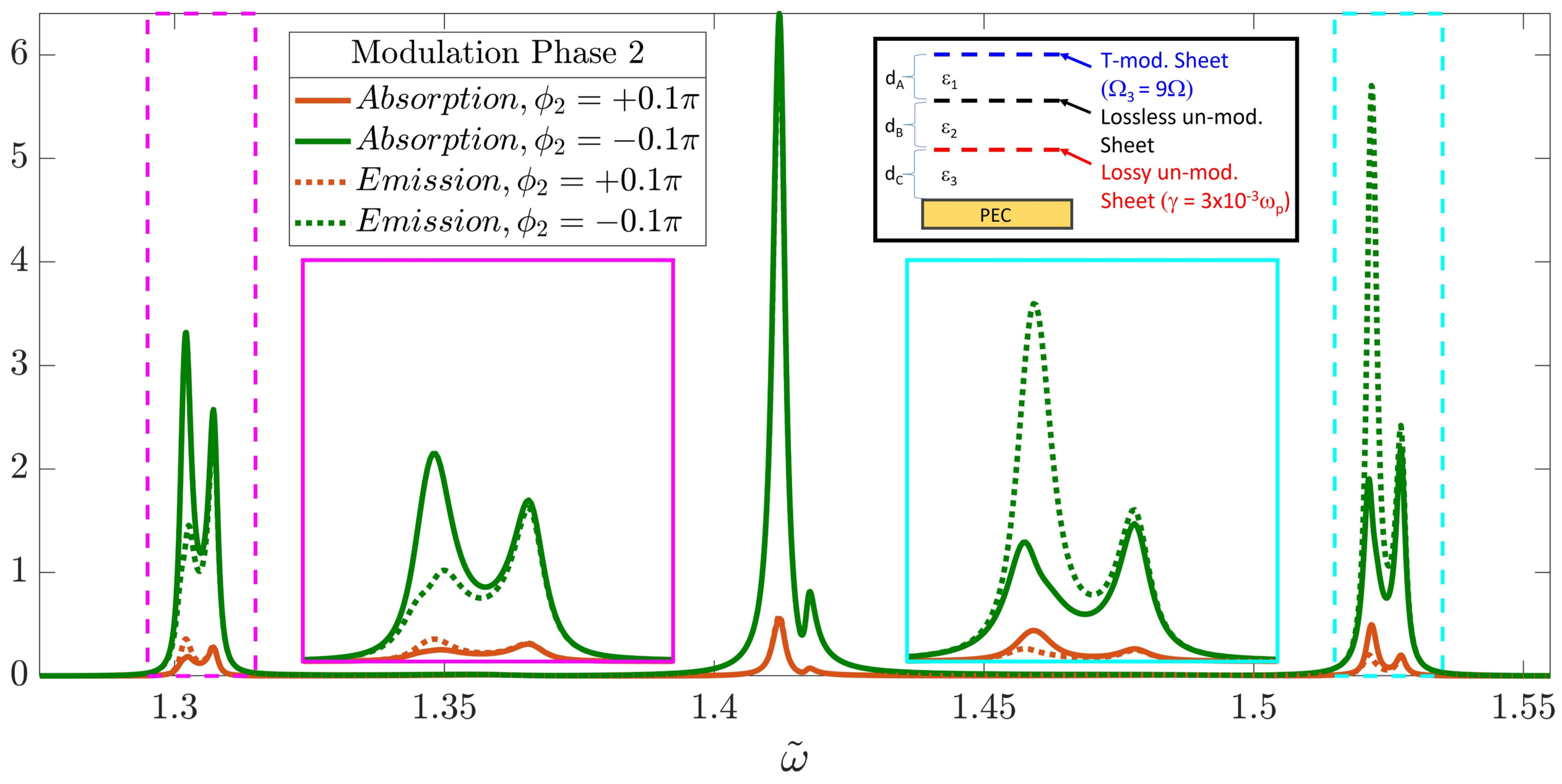

***Figure 4. Large violation of Kirchhoff's law of thermal radiation in the phased Floquet metasurface system.*** Absorption (solid lines) and emission (dotted lines) spectra are shown for two modulation phases, $\phi_2 = +0.1\pi$ (orange) and $\phi_2 = -0.1\pi$ (green), revealing asymmetric absorptivity and emissivity that violate Kirchhoff's thermal radiation law. The insets provide blow-ups in spectral regions with strong asymmetry, marked by magenta and cyan dashed boxes. The schematic in the top right corner depicts the layered structure of the Floquet sheets, including the time-modulated sheet $\Omega_3 = 9\tilde{\Omega}$, the lossless unmodulated sheet, and the lossy unmodulated sheet ($\gamma = 3\times10^{-3}\omega_p$), over a ground plane.

# Supplemental information

## Note S1: Robustness check of the modulation phase and the geometric parameters

To evaluate the stability and experimental feasibility of the proposed nonreciprocal mechanism, we investigate the system's robustness against variations in two critical parameters: the modulation phase $\phi_2$ and the interlayer spacing $d_{\mathrm{B}}$. Robustness of the Modulation Phase. First, we examine the modulation phase $\phi_2$ . While the manuscript presents the optimal asymmetric frequency conversion at $\phi_2 = 0.1\pi$, we simulated the optical response across a broad range of phase values spanning from $\phi_2 = 0.075\pi$ to $0.25\pi$. As shown in Figure S1, although the conversion efficiencies and peaks exhibit expected variations as the phase deviates from the optimal value, the key asymmetrical feature remains robust across the desired frequency range. This confirms that the proposed effect is stable as the phase condition deviates from the optimal value.

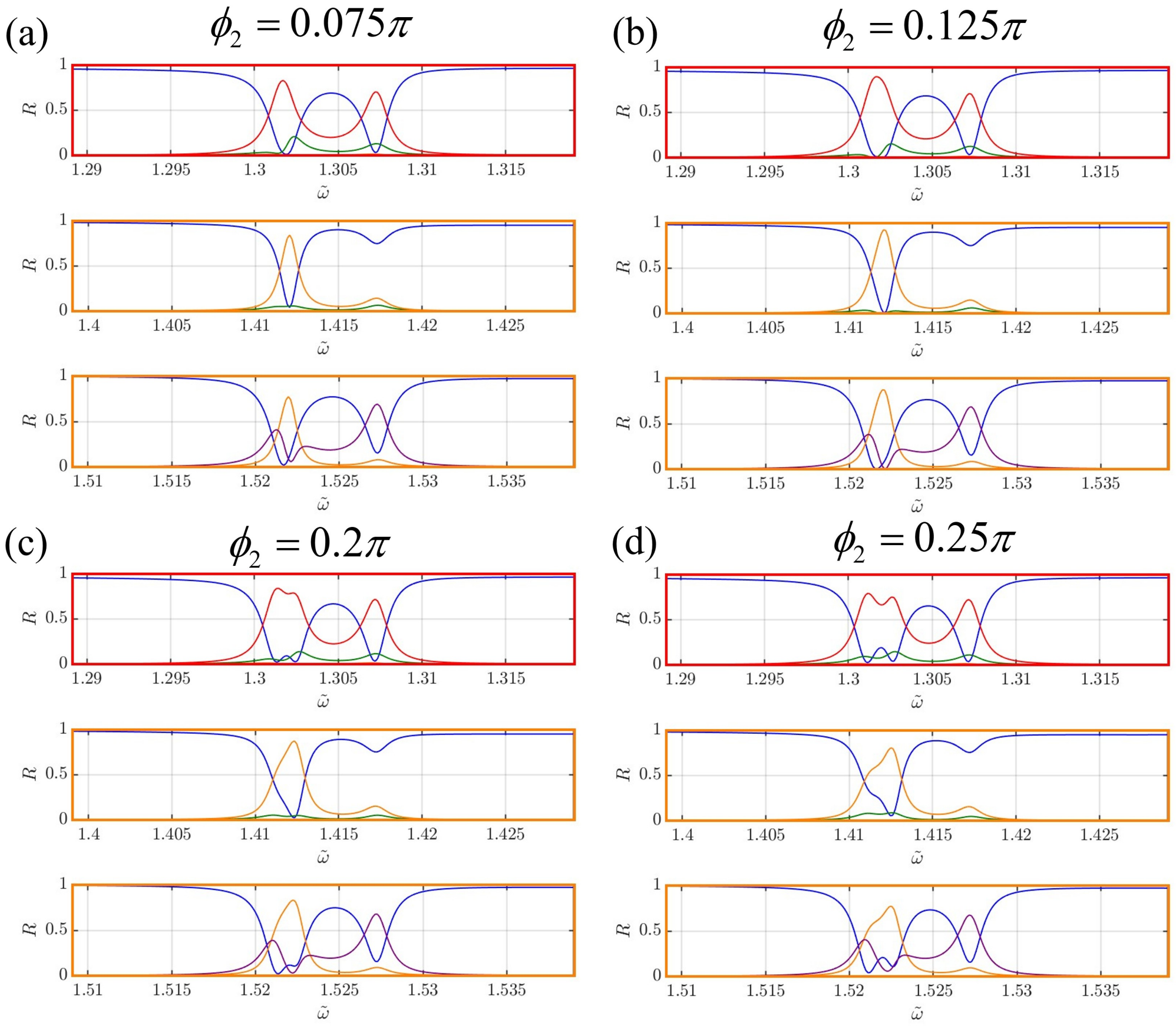

Figure S1. Robustness of the asymmetric frequency conversion against detuning of the modulation phase, related to Figure 3. The optical response spectra are calculated for (a) $\phi_2 = 0.075\pi$, (b) $\phi_2 = 0.125\pi$, (c) $\phi_2 = 0.2\pi$, and (d) $\phi_2 = 0.25\pi$.

Next, we assess the impact of geometric detuning by varying the parameter $d_\mathrm{B}$, which directly governs the resulting modal spacing of the surface modes. The ideal equal-spacing condition, which occurs at $d_\mathrm{B}$ = 2.7 in our normalized units, is designed to perfectly phase-match the intermodal transitions, thereby maximizing the frequency conversion efficiency. However, further analysis reveals that this is not a strict condition for the nonreciprocal phenomenon. For the detuned cases ranging from $d_\mathrm{B}$ = 2.65 to $d_\mathrm{B}$ = 2.73 (see Figure S2), the conversion peaks experience slight frequency shifts and minor efficiency fluctuations; nevertheless, they remain clearly observable, indicating fair structural tolerance. From an experimental fabrication perspective, the large geometric error tolerance translates well to the system fabrication. If the proposed multi-sheet structure is implemented in the microwave regime, the dimensions scale up to the centimeter range, and the detuning range ( $\Delta d_B \approx \pm 0.03$ ) corresponds to a fabrication tolerance of approximately 1 millimeter. Given standard PCB making and mechanical machining, the proposed system is robust against practical manufacturing errors.

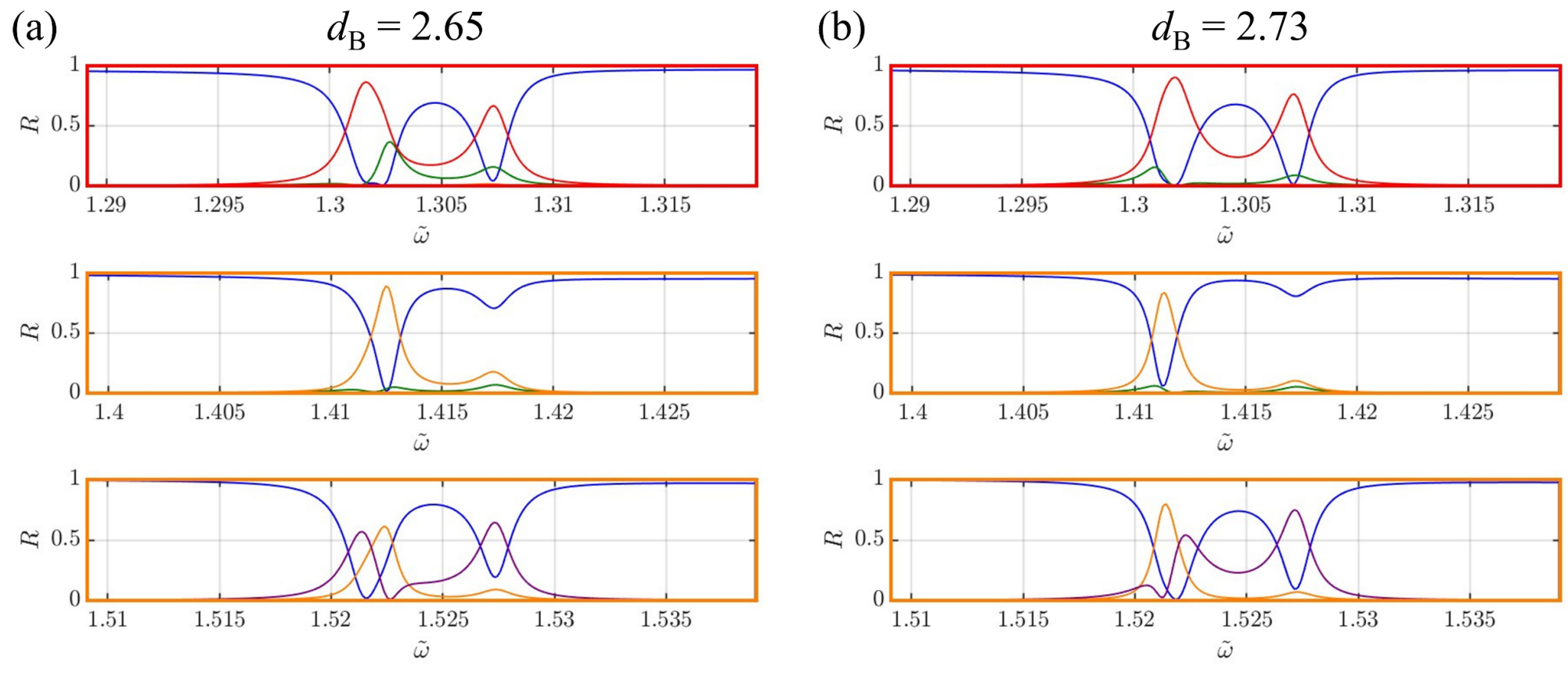

Figure S2. Robustness of the asymmetric frequency conversion against detuning of the interlayer spacing, related to Figure 3. Asymmetric frequency conversion spectra for a detuned interlayer spacing (a) $d_B = 2.65$ and (b) $d_B = 2.73$. The modulation phase is set to $\phi_2 = 0.1\pi$.

# Note S2: COMSOL simulation setup

Full-wave simulations are performed using the RF Module in COMSOL Multiphysics. To model the frequency conversion induced by the multi-tone temporal modulation ($\Omega$, $2\Omega$, $9\Omega$), we employed a frequency-domain multi-physics coupling approach:

- Multi-Channel Coupling: We coupled 27 "Electromagnetic Waves, Frequency Domain" (emw) physical interfaces in COMSOL. Each n-th interface is shifted to operate at $f \pm n\Omega / 2\pi$ (for integer $n \in [-13, 13]$) via the Equation View, where *f* is the frequency of the incident TM plane wave.
- Time-Modulated Sheet: The modulated top sheet is defined as an infinitely thin boundary using the "Surface Current Density" node. To capture the generation of sidebands, the current density for the n-th channel is explicitly defined as:

$$J_{x,n} = \frac{W_{p0}}{\gamma_i - i\omega_n}\begin{pmatrix} E_{x,n} + \alpha_1 e^{-i\phi_1} E_{x,n-1} + \alpha_1 e^{i\phi_1} E_{x,n+1} + \alpha_2 e^{-i\phi_2} E_{x,n-2} + \alpha_2 e^{i\phi_2} E_{x,n+2} \\ +\alpha_3 e^{-i\phi_3} E_{x,n-9} + \alpha_3 e^{i\phi_3} E_{x,n+9} \end{pmatrix}$$

  where $W_{p0} / \left(\gamma_i - i\omega_n\right)$ is the complex conductivity, and $n \in \{1,2,9\}$. The bottom two sheets are modeled similarly but without the modulation terms.
- Boundaries & Manley–Rowe Relations: The lateral boundaries are set to Floquet Periodic Conditions $k_x = 0.6$, while the top and bottom are truncated with Perfectly Matched Layers (PMLs) and a PEC ground, respectively. To satisfy the Manley–Rowe relations, the calculated reflectance $R_n$ for each harmonic channel is obtained via the ratio of the reflected time-averaged Poynting vectors to the incident one, and then is multiplied by the frequency factor from Manley–Rowe relations.

- In Figure S3, we depict the schematic for our COMSOL setup. To simulate an obliquely incident plane wave in an infinite planar structure, below the PML, an input port is assigned to launch the incident wave. A field-collecting line is also placed to extract the scattered reflection harmonics rather than the 0th order reflection. The simulation domain consists of three parallel conductive sheets in air: a time-modulated top sheet (red), an unmodulated lossless middle sheet (green), and an unmodulated lossy bottom sheet (blue). The normalized interlayer spacings are configured as $d_A = 1$ and $d_B = 2.7$, while the distance from the bottom sheet to the underlying PEC reflector is set to $d_C = 20$.

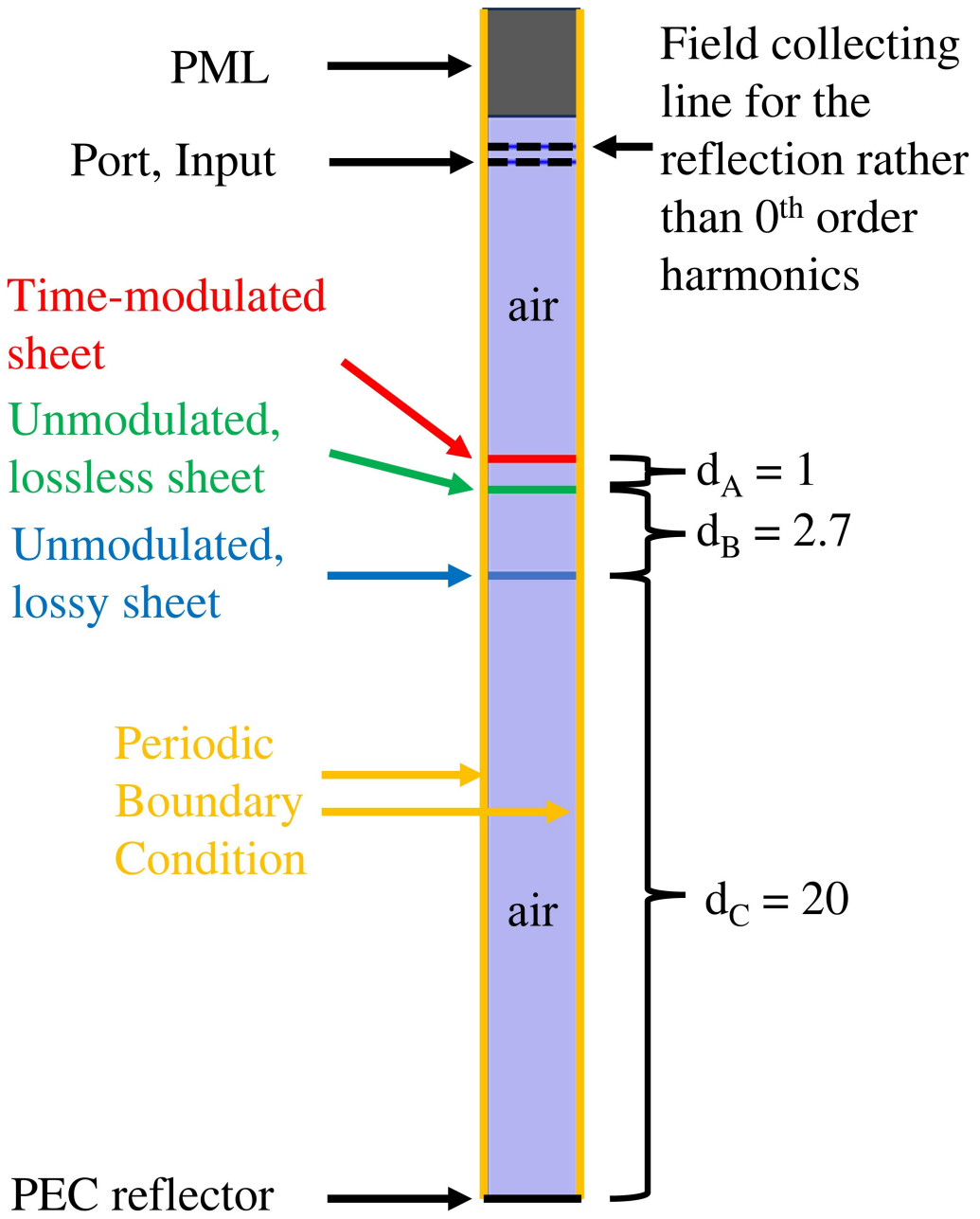

Figure S3. Schematic of the COMSOL simulation domain, related to Figure 3.